\documentclass{aastex631}

\begin{document}

\title{Bulk Strong Matter: the Trinity}

\author[0000-0002-3093-8476]{Xiaoyu Lai}
\affiliation{Department of Physics and Astronomy, Hubei University of Education, Wuhan 430205, China}
\affiliation{Research Center for Astronomy, Hubei University of Education, Wuhan 430205, China}

\author[0000-0002-3388-1137]{Chengjun Xia}
\affiliation{Center for Gravitation and Cosmology, College of Physical Science and Technology, Yangzhou University, Yangzhou 225009, China}

\author[0000-0002-9042-3044]{Renxin Xu}
\affiliation{School of Physics, Peking University, Beijing 100871, China}
\affiliation{Kavli Institute for Astronomy and Astrophysics, Peking University, Beijing 100871, China}

\begin{abstract}

Our world is wonderful because of the normal but negligibly small baryonic part (i.e., atoms) although unknown dark matter and dark energy dominate the Universe.
A stable atomic nucleus could be simply termed as ``strong matter'' since its nature is dominated by the fundamental strong interaction.
Is there any other form of strong matter?
Although nuclei are composed of 2-flavoured (i.e., up and down flavours of valence quarks) nucleons, it is conjectured that bulk strong matter could be 3-flavoured (with additional strange quarks) if the baryon number exceeds the critical value, $A_{\rm c}$, in which case quarks could be either free (so-called strange quark matter) or localized (in strangeons, coined by combining ``strange nucleon'').
Bulk strong matter could be manifested in the form of compact stars, cosmic rays, and even dark matter.
This trinity will be explained in this brief review, that may impact dramatically on today's physics, particularly in the era of multi-messenger astronomy after the discovery of gravitational wave.

\end{abstract}


\section{Bulk strong matter: An introduction} \label{sec:intro}

Atoms/molecules are the building units of normal matter, either condensed or fluid, to be manifested in the form of solid, liquid and gas.
There is no essential difference if atoms are ionized significantly, so-called plasma, except for the electromagnetic interaction at a longer distance (besides the short-distance collisions between units).
Certainly, the state of matter above is primarily determined by the electromagnetic interaction.
Can a kind of matter exist, the state of which is controlled by the fundamental strong interaction?
Yes, atomic nucleus is such an example, in which nuclear force dominates.
Nevertheless, in this report, we would like to discuss two kinds more: cold quark matter and strangeon matter; the basic unit of the former is by definition quarks while that of the latter is strangeons.

It is well known that the strong and the electromagnetic interactions are two of the four fundamental interactions that dictate our world.
The strong one not only confines quarks inside nucleons, but also holds nucleons together to form atomic nuclei.
Atoms that composed of nuclei and electrons are the building blocks of ordinary matter, which could be named ``electromagnetic matter'' (or simply ``electric matter'') since its properties are actually dominated by the electromagnetic interaction.
Similarly, a stable atomic nucleus, though microscopic, could then be termed as ``strong matter'' since its major nature is controlled by the strong interaction.
As the interaction is strong and of short range, the density ($\sim 10^{15}$ g/cm$^3$) should be much higher than that of electric matter ($\sim 10$ g/cm$^3$).

In this article, we focus on the so-called ``bulk strong matter'', which could be macroscopic, with a size large enough (i.e. large baryon number) so that the surface energy is negligible compared to the bulk energy.
Microscopic strong matter, i.e., normal nuclei in atoms, is 2-flavour symmetric (or isospin symmetric) with comparable amounts of protons and neutrons, and this is the reason that an atomic nucleus is electrically positive.
For the microscopic strong matter, electrons would emerge with approximately half the amount of baryons.
In addition, due to the smallness of an atomic nucleus and the relatively weak electromagnetism-force, electrons generally locate outside the nucleus.
Well, how is the cases for bulk strong matter (i.e., macroscopic strong matter)?

Besides the strong interaction, the weak one is also essential to determine the nature of bulk strong matter.
In view of the importance of electrons participating in the weak interaction, we may use the electron Compton wavelength, $\lambda_{\rm c} = h/(m_{\rm e}c)$, to mark the boundary between micro- and macro- strong matter, and the critical baryon number, $A_{\rm c}$, could thus be order of $\lambda_{\rm c}^3/{\rm fm^3} \sim 10^9$, where the average volume of one baryon in strong matter is supposed to be $\sim $fm$^3$.
If we pile up nucleons (including neutrons and protons) to make a gigantic nucleus, electrons will become almost inside the nucleus and be extremely energetic (up to $\sim 300$ MeV), therefore a giant nucleus could hardly keep stable due to those energetic electrons.
As the baryon number increases, if its constitutions are still neutrons and protons, the growing nucleus will become more and more neutron-rich, via the weak interaction of $e^-+p \rightarrow n+\nu_e$.
Actually, in 1932, an idea of gigantic nucleus was conceived by Landau~\citep{Landau1932} for the first time, which is then developed to be the very elaborated models of conventional neutron stars, especially after the discovery of radio pulsars.
However, in this paper we would only like to discuss the bulk strong matter which is stable at zero pressure, as in the case of atomic nucleus.
By this definition, neutron matter would not be included as a form of bulk strong matter, since pressure-free neutron matter is unstable against the weak decay of $n \rightarrow p+e^-+{\bar \nu}_e$ in vacuum or low-density environment.

Nonetheless, the macroscopic strong matter formed by compressing a huge number of normal nuclei, on the other hand, may not still be 2-flavoured since the fundamental weak interaction can change quark flavours not limited only to the up and down.
The inclusion of the third flavour, strangeness (strange quark, light-flavoured too), to stabilize the pressure-free bulk strong matter has attracted particular attentions.
3-flavour (up, down and strange quarks) symmetry could be restored when normal matter becomes so dense that 2-flavoured nuclei come in close contact and eventually forms bulk strong matter.
As will be demonstrated in this article, the restoration of 3-flavour symmetry could be essential when 2-flavoured normal matter is squeezed and forms bulk strong matter during a core-collapse supernova.
Moreover, in this article we are addressing that quarks could probably be localized in clusters (i.e., strangeons), rather than free, in 3-flavour symmetric strong matter, which serves as an alternative solution to the equation of state (EOS) of stellar matter at supra-nuclear density.
This idea is truly in the regime of ``old physics'', but may have additionally consequences for us to understand dark matter and even ultra-high energy cosmic rays.

\begin{figure}[ht!]
\centering
\includegraphics[width=0.8\linewidth]{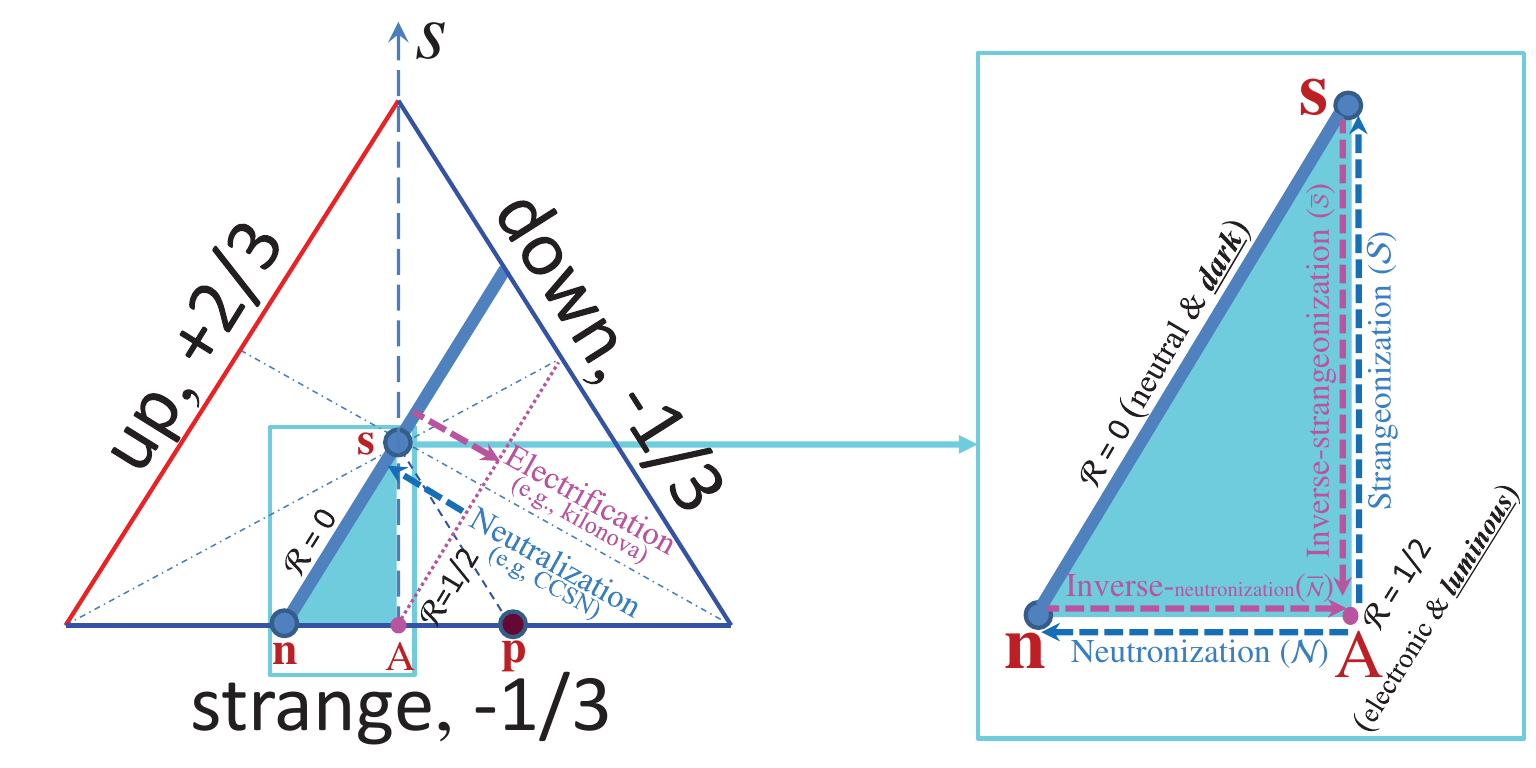}
\caption{An updated triangle of light-quark flavours~\citep{Xu2021}. The points inside the triangle define the states with certain quark number densities of three flavours (\{$n_{\rm u}, n_{\rm d}, n_{\rm s}$\} for up, down and strange quarks), measured by the heights of one point to one of the triangle edges.
Axis $S$ denotes strangeness, with perfect isospin symmetry.
Normal nuclei are around point ``A'', conventional neutron stars are in point ``n'', extremely unstable proton stars are in point ``p'', and strange stars (both strange quark star and strangeon star) in point ``s'' in the center of the triangle with $n_{\rm u}=n_{\rm d}=n_{\rm s}$.
The charge-mass-ratio of quarks at point ``A'' is $\mathcal{R}\simeq 1/2$, but $\mathcal{R}=0$ on the line of ``ns''; the former is ``electronic'' and ``luminous'', while the latter could hardly be detected because of neutrality and would take a ``dark'' role as matter.
The zoomed $\triangle$snA shown in the right indicates that, supernovae explosions (from ``A'' to a piont in line ``ns'') realize neutronization ($\mathcal{N}$) or strangeonization ($\mathcal{S}$), and kilonovae (from a piont in line ``ns'' back to ``A'') realize inverse-neutronization ($\bar{\mathcal{N}}$) or inverse-strangeonization ($\bar{\mathcal{S}}$) (see \S\ref{sec:kilonova}).
\label{fig:triangle}}
\end{figure}

In conclusion, we may summarize the concepts discussed above (atomic nucleus, neutron star, strange quark star and strangeon star) in a triangle illustrated in Fig.~\ref{fig:triangle}.
Because of baryon conservation, it is convenient to discuss the quark numbers of the three flavours in the triangle, for a given baryon density, $n_{\rm b}=(n_{\rm u}+n_{\rm d}+n_{\rm s})/3$, with quark number densities of up ($n_{\rm u}$), down ($n_{\rm d}$) and strange ($n_{\rm s}$).
It is shown that the bottom strange edge is divided into three equal parts by points ``n'' and ``p''  because the triangle ``$\triangle$snp'' is left-right symmetrical to the ``$S$''-axis but shrinks by two-thirds.
Normal nuclei are around point ``A'', onventional pure neutron stars are in point ``n'', extremely unstable proton stars are in point ``p'', and strange stars (both strange quark star and strangeon star) in point ``s''.
It is evident that the flavour symmetry could be the key to understand the nature of bulk strong matter.
In light of the challenging problem in understanding dense matter and compact star, this flavour symmetry could be an issue, to some sense, in response to the spirit of ``{\em Symmetry dictates interaction}''~\citep{YangCN1980}.

Bulk strong matter could be manifested in the form of compact stars, cosmic rays, and dark matter.
This trinity will be explained in this article.

\section{The nature of bulk strong matter}

\subsection{Quark matter: 2- or 3-flavoured?}

Despite that there are six flavours of quarks, we usually neglect heavy flavours of quarks (charm, top and bottom quarks) in the study of dense matter at a few times of the nuclear saturation density, due to the energy scale, $E_{\rm scale}$.
Estimated with the Heidelberg's relation, the energy scale would be order of $E_{\rm scale}\sim \hbar c/\Delta x \sim 0.5 {\rm GeV} < m_{\rm heavy}$, where the separation between quarks is $\Delta x \sim 0.5$ fm and $m_{\rm heavy}$ is the mass scale of heavy flavours of quarks.
This energy scale, however, is much higher than the mass difference between strange and up/down quarks ($\Delta m_{\rm uds} \sim 0.1$ GeV), $E_{\rm scale} \gg \Delta m_{\rm uds}$.
Therefore, we may expect 3-flavoured matter at a few times of the nuclear saturation density.
But what is the exact form of dense matter with three flavours of light quarks?

Strange quark matter, composed of comparable amounts of up, down and strange quarks, could be more stable than ud quark matter, as proposed by Witten and many others~\citep{Bodmer1971,Witten1984}.
Although it remains uncertified due to the difficulty in non-perturbative quantum chromo-dynamics (QCD), this conjecture seems reasonable, because the additional strange flavour provides an extra degree of freedom to lower the Fermi energy in the free-quark approximation.

A general and extendable way to illustrate the so-called Witten's conjecture is from the concept of symmetry, explicitly the light-flavour (3-flavour) symmetry.
The restoration of light-flavour symmetry ensures neutrality without electrons.
This is not the case for 2-flavoured ($u$ and $d$ quarks) quark matter, because the 2-flavour symmetric strong matter is positively charged and requires electrons to emerge with a number about half the baryon number to ensure neutrality.
These electrons in bulk strong matter are relativistic and energetic enough to make it $d$-rich, i.e. the state with 2-flavour {\em asymmetry}, which is the case for neutron matter.
It is interesting to note that these electrons become insignificant for a micro-nucleus (with 2-flavour symmetry) because it is too small (with baryon number smaller than $A_{\rm c}$) in size so that all the electrons are outside and non-relativistic.

Certainly, the Witten's conjecture is based on the MIT bag model with negligible interaction between quarks, although it can be illustrated by the ``light-flavour maximization'' as above.
As shown in \S 2.2, the interaction between quarks is non negligible; however, taking into account the real non-perturbative interaction is a difficult task.
Under some phenomenological models for interacting quarks, the quark matter composed of up and down quarks ($ud$QM) could be more stable than strange quark matter~\citep{Holdom2017}, and the astrophysical consequences of compact stars composed of $ud$QM have discussed combined with recent observations~\citep{Zhang2020}.

The binding energy of the system should be calculated from first QCD-principles, but this is actually unlikely because of the rich non-perturbative effects.
One then has to either model phenomenologically or conjecture from view-point of symmetry.
Among all the candidate models, in this paper we explore the possibility of strong matter made of strangeons, which follows the light flavor-symmetry.
This can be attributed to the SU(3) flavor-symmetry in hadron spectroscopy, where slight flavor symmetry breaking may persists as strange quarks are more massive.
Similarly, the total number of strange quarks may be slightly less in strangeon matter according to our model calculation~\citep{Miao2022IJMP}.
It is worth mentioning that, the electrons contribution to the energy per baryon is included in~\citet{Miao2022IJMP} but note in~\citet{Holdom2017}.

The study of $ud$QM is certainly interesting, showing that the state of dense matter has various possibilities in the theoretical points of view.
But many other possibilities have also been announced.
%
%
In the \S 2.2, we conjecture a state composed of strangeons which are also the result of significant interaction between quarks.

\subsection{Strangeon matter}

The perturbative QCD, based on asymptotic freedom, works well only at high energy scales, $E_{\rm scale}>\Lambda_{\rm QCD} \sim 1$ GeV.
However, the state of pressure-free strong matter at supra-nuclear density should be relevant to non-perturbative QCD because $E_{\rm scale} < \Lambda_{\rm QCD}$, which is exactly a similar case of normal atomic nuclei.
A conjecture of ``condensation'' in position space, with strange quark cluster as the constituent units~\citep{Xu2003}, rather than condensation in momentum space for a color super-conducting state, was thus made for cold matter at supra-nuclear density.
The strange quark cluster is renamed strangeon, being coined by combining ``strange nucleon" for the sake of simplicity~\citep{XG2016,Wang2017ApJ}.

Starting from deconfined quark state with the inclusion of strong interaction between quarks, and using the Dyson-Schwinger-Equation approach to the non-perturbative QCD~\citep{Fischer2006}, one would estimate that the strong coupling constant $\alpha_{\rm s}$ could be greater than 1 at density $\sim 3 \rho_0$.
This means that a weakly coupling treatment is inadequate for realistic dense matter in compact stars.
It is also worth noting that a weakly coupling strength comparable with that of quantum electrodynamics ($1/137<0.01$) is possible only if the baryon number density is unbelievably and unrealistically high ($n_{\rm B}>10^{123}n_0$).
From this point of view, although some efforts have been made to understand the state of pulsar-like compact stars in the framework of conventional quark stars, including the MIT bag model with almost free quarks~\citep{Alcock1986} and the color-superconductivity state model~\citep{Alford2008}, realistic densities inside pulsar-like compact stars cannot be high enough to justify the validity of perturbative QCD.
The strong coupling between quarks may naturally render quarks grouped in quark-clusters, and each quark-cluster/strangeon is composed of several quarks condensating in position space rather than in momentum space.

Certainly, whether quarks would be grouped in quark-clusters is hard to answer from direct QCD calculations.
We have attempted to answer this question by making an analog between quark-clusters and water molecules~\citep{Xu2009csqcd}, where the latter are also clusters made of electrons, protons and oxygen nuclei.
The much weaker electromagnetic coupling can group particles into clusters, so we state that the strong coupling between quarks may also ``naturally" render quarks grouped in quark-clusters.
The astrophysical compressed baryonic matter, therefore, could be in a state of strangeon matter.

In principle, nucleons can also be viewed as clusters made of three valence quarks.
Since the mass of nucleons is small enough with significant quantum effects, nucleons form a liquid phase of nuclear matter.
However, at smaller densities ($\lesssim 0.08 {\rm fm}^{-3}$), the clustering effects take place and form large nuclei, where various types of crystalline structures are formed.
For the cases of strangeon that comprised of a large number of quarks, the mass is much larger and the formation of solid structures become more likely because of short quantum wavelength $\sim h/(mc)$.
The melting temperature would be much larger than the temperature inside pulsars~\citep{Xu2003}.

Although the state of bulk strong matter is essentially a non-perturbative QCD problem and is difficult to answer from first principles, the astrophysical point of view could give some hints that bulk strong matter could be in the form of strangeon matter.
Strangeon matter may constitutes the true ground state of strongly-interacting matter rather than $^{56}$Fe and neutron matter, and this could be seen as a {\it generalized Witten's conjecture}.
The {\it traditional Witten's conjecture} focus on the matter composed of almost free $u$, $d$ and $s$ quarks, and we generalized it to the statement that ``strange matter composed of $u$, $d$ and $s$ quarks, which are free or localized, could be more stable than $^{56}$Fe''. Certainly, this is distinctively different from (but related to) Witten's conjecture since quarks may still be confined in the generalized Witten's conjecture.

Strangeon matter is conjectured to be the compressed baryonic matter of compact stars, where strangeons form due to both the strong and weak interactions and become the dominant components inside those stars.
Considering the similarity between strangeons and nucleons, we used a general and phenomenological model, the Lennard-Jones model~\citep{LX2009b}, to describe the EOS of strangeon stars.
Although the Lennerd-Jones potential originally describes the interaction between inert gas molecules, it has the character of long-range attraction and short-range repulsion which is also the character of the interaction between nucleons.
Moreover, the Lennerd-Jones model also shows the similarity between strangeon and inert gas molecule that, the former  is colorless and the latter is chargeless.

Then what could be a realistic strangeon?
It is known that $\Lambda$ particles (with structure $uds$) possess light-flavor symmetry, and one may think that a kind of strangeons would be $\Lambda$-like.
However, the interaction between $\Lambda$'s could be attractive, which would render more quarks grouped together.
Motivated by recent QCD simulations of the $H$-dibaryons (with structure $uuddss$), a possible kind of strangeons, $H$-clusters, is proposed~\citep{LGX2013}.
Moreover, because of their classical behavior (as a large mass may result in a small quantum wavelength), strangeons that exist in compact stars could locate in periodic lattices (i.e., in a solid state) when temperature becomes sufficiently low.
Certainly, strangeon stars with global rigidity would have broad astrophysical interests and have significant implications in astronomical observations, as discussed in \S 3.1.

\begin{figure}[ht!]
\centering
\includegraphics[width=0.6\linewidth]{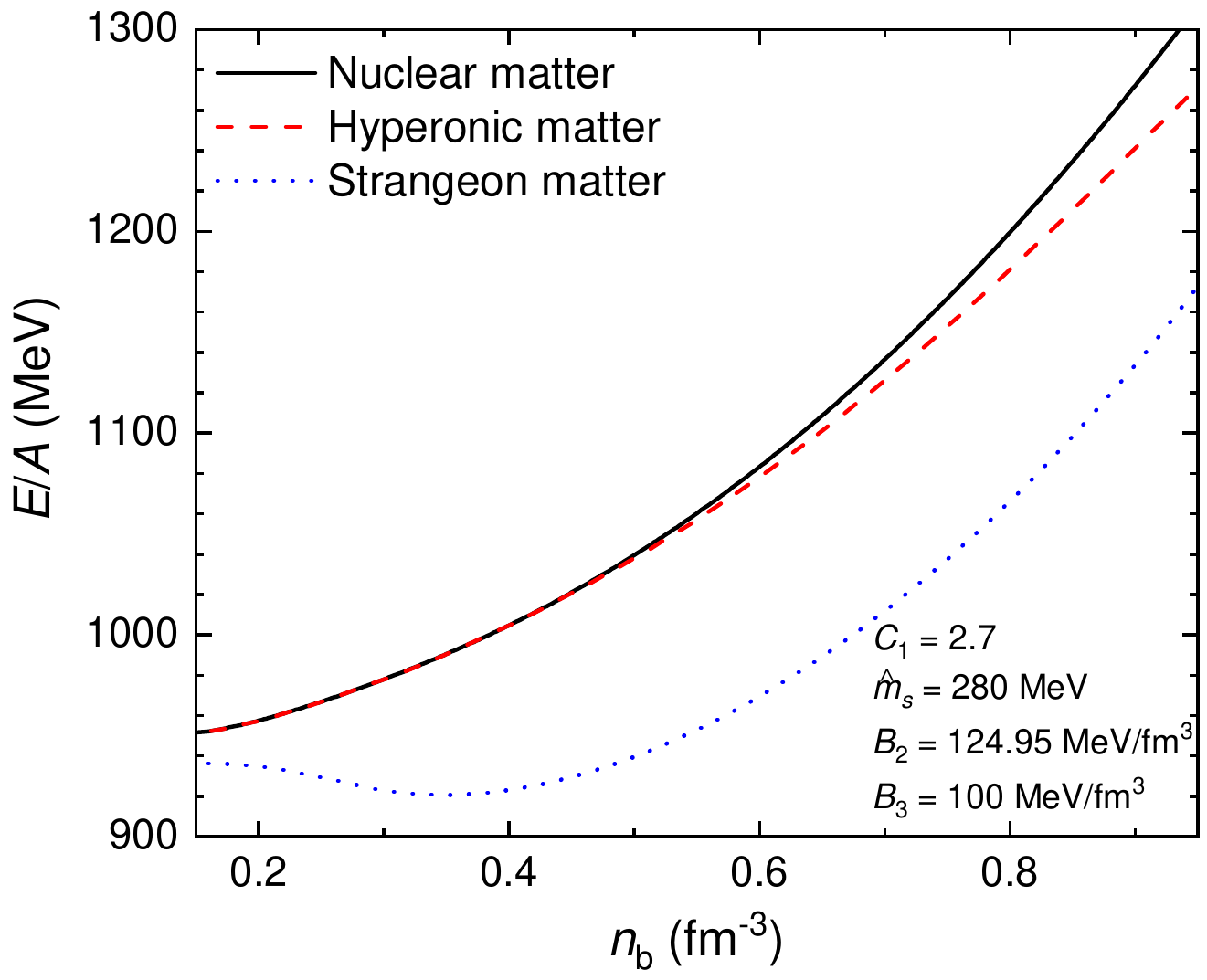}
\caption{Energy per baryon of nuclear matter (dash-dotted), hyperonic matter (dashed), and strangeon matter (solid) obtained in the framework of a linked bag model, where four different parameter sets are adopted (see the details in~\citep{Miao2022IJMP}).\label{fig:epb}}
\end{figure}

In fact, to gain some insight on the stability of strangeon matter, recently we proposed a linked bag model~\citep{Miao2022IJMP}. The bags containing $N_q$ (= 3, 6, 9, ...) quarks form a simple cubic lattice, where the interaction between two or more bags is treated by physically connecting the bags. The energy per unit cell is then fixed by
\begin{equation}
\label{eq:energy per cell}
E=\sum_{i=u,d,s,e}(\Omega_i+N_i\mu_i)+B a^3-\frac{z_0}{r_{\rm bag}}\frac{\omega}{4\pi},
\end{equation}
where $\Omega_i$, $N_i$ and $\mu_i$ represent the thermodynamic potential, total particle number, and chemical potential of particle type $i$ (including quarks and electrons).
$B$ is the bag parameter, V is the enclosed volume of the bag, and the variable $\omega$ represents the solid angle
of the remaining bag.
The model parameters are firstly calibrated to reproduce the saturation properties of nuclear matter, and the derived values are then applied to strangeon matter to obtain the corresponding energy per unit cell with Eq.~(\ref{eq:energy per cell}).
As an example, in Fig.~\ref{fig:epb} we present the energy per baryon for nuclear matter, hyperonic matter, and strangeon matter in compact stars using four different parameter sets. It is evident that the minimum energy per baryon of strangeon matter can be smaller than than $^{56}$Fe, i.e., fulfilling the {\it generalized Witten's conjecture}. Meanwhile, the corresponding density is approximately twice the nuclear saturation density and the EOS becomes stiffer than nuclear matter scenarios.

\subsection{Trinity of bulk strong matter}

We have addressed that, due to the strong coupling between quarks in the compressed baryonic matter in astrophysics, quarks may be naturally grouped in strangeons, and each strangeon is composed of several quarks condensating in position space rather than in momentum space.
Strangeon matter is the form of bulk strong matter.
Then what are the consequences of this strong matter?
In the regime of free quarks, Witten conjectured an absolutely stable state of strange quark matter~\citep{Witten1984}, and addressed dramatic consequences of this strong matter: quark star produced during supernova, residual quark nuggets after cosmological QCD phase transition, as well as strange cosmic rays.
These three are retained if quarks are not free but localized in strangeons, and strangeon matter shares a similar trinity: compact stars, dark matter, and cosmic rays.

\section{Compact stars}

Compact stars composed totally of strangeons are strongeon stars.
It is proposed that, as we have stated above, the pulsar-like compact stars could actually be ``strangeon stars'' composed totally of strangeons, and this proposal is also motivated by several astrophysical points of view~\citep{Xu2003}.
Being similar to strange quark stars, strangeon stars have almost the same composition from the center to the surface.
Strangon stars are self-bound by the residual interaction between strangeons, while strange quark stars by bag-like confinement.
The in-between interaction of strangeons could be regarded as a natural extension of the nuclear force between nucleons, simply with different flavour numbers of quarks, from 2 to 3.
The properties of strangeons, as well as the strangeon matter surface, could then be calculated, to be helpful for understanding different observations of pulsar-like compact stars (see the review~\cite{LX2017} and references therein).
Significant attention has been paid to the peculiar observational features related to both the surface condition and the global structure of strangeon stars.

\subsection{Surface properties}

Similar to quark stars, strangeon stars are self-bound by strong interaction on surface.
Some problems related to the surface properties of pulsar-like compact stars can be naturally solved in bare strangeon star scenario.
(1) A bare strangeon star does not have atomic features in the thermal X-ray emission because no atom exists on its surface.
(2) A strangeon matter surface could naturally explain the pulsar magnetospheric activity~\citep{Xu1999, YX2011} as well as the subpulse-drifting of radio pulsars~\citep{Lu2019}.
(3) The strong radiation pressure caused by thermal emission from strangeon star might play an important role in promoting core-collapse supernovae~\citep{Chen2007} and even in explaining long-lived plateau of light curves of GRB afterglow~\citep{Dai2011}.
The neutrino burst observed during supernova 1987A could also be understood in the regime of strangeon star~\citep{Yuan2017}.
(4) 
The motion of the electron sea on the magnetized surface was investigated~\citep{Xu2012}, and it is found that the absorption feature of 1E 1207.4-5209 could be understood in this hydrocyclotron oscillations model.
(5) The strangeness barrier could help us to understand the redshifted O VIII Ly-$\alpha$ emission line and the change in the blackbody radiation area of 4U 1700+24~\citep{Xu2014RAA}, and the plasma atmosphere of strangeon stars can reproduce the Optical/UV excess observed in X-ray dim isolated neutron stars~\citep{Wang2017ApJ}.

It is worth noting that the peculiar surface difference between strangeon star and normal neutron star would have profound implication to the magnetospheric activity related to coherent radio emission of compact stars.
Besides trying to alleviate the binding energy problem in pulsar radio  emission~\citep{XuQiao1998}, one could also find a reliable way to create a bulk of large bunches with energetic $e^\pm$-particles just above stellar surface in the strangeon star model, because of the solid and rigidity nature of strangeon matter.
The rough solid surface may help triggering a lot of bunches of electron-positron pairs, that would work both for normal radio pulsars radiating regular pulses and for occasional events sending irregular bursts~\citep[e.g., repeating fast radio bursts, FRB;][]{WangWY2022,WangWY2021} .
In the future, studies combining radio pulsar with FRB are welcome in order to understand the electrodynamics and related radiative mechanisms of rotating compact stars with strong magnetic fields.

In addition to the surface features, the global properties, especially the maximum mass and the tidally deformability, are more closely related to the EOS.
Pulsar glitch is a useful probes into pulsars' interior structure, and some observational properties of glitches are studied in strangeon star model.
The multiwavelength electromagnetic counterparts of GW170817, named kilonova, can also give constraints on the nature of compact stars, so it is necessary to explore the kilonova-like event associated with merging strangeon stars.
These topics will be reviewed in the following.

\subsection{Stiff equation of state and massive strangeon star}

Some properties of strangeon stars are similar to that of quark stars because of the self-bound nature.
Nevertheless, there are unique properties related to the clustered and solid structure, as we will explained in this and the next subsections.

The strangeon star model predicts high mass pulsars~\citep{LX2009b,LX2009a} before the discovery of pulsars with $M > 2 M_\odot$~\citep{Demorest2010}.
The EOS of strangeon matter would be stiffer than that of nuclear matter for two main reasons: (1) strangeons should be non-relativistic particles because of their large mass, and (2) there could be strong short-range repulsion between strangeons (an analogy for the repulsive core of nuclear force).

Experimentally, interaction between nucleons and that between atoms/molecules share the common nature of long range attraction and short range repulsion, although the former is strong interaction and the latter is electromagnetic interaction.
In view of the fact that atoms/molecules are chargeless and both nucleons are colorless, we may expect the interction between strangeons also exhibits the nature of long range attraction and short range repulsion, because strangeons as the strong units are colorless.

We further assume that the Lennard-Jones model, which applies to the interaction between molecules, also applies to the strangeons.
Besides the very low masses ($< 10^{-2} M_{\odot}$) a strangeon star can has as a direct consequence of self-bound surface, we found that, under the Lennard-Jones-like inter-strangeon interaction, strangeon stars could have high maximum masses ($>2M_{\odot}$) because of the stiff EOS~\citep{LX2009b}.
Later, the stiff EOS was indicated observationally by radio observations of PSR J1614-2230, which implied that the pulsar mass is $1.97\pm 0.04 M_{\odot}$~\citep{Demorest2010}.
It is worth emphasizing that, although the state of dense matter is conventionally thought to be soft and thus cannot result in high maximum mass if pulsars are quark stars, strangeon stars model predict massive pulsars before the observational discovery.
Moreover, the observations of pulsars with higher mass (e.g. $>2.5 M_{\odot}$) would even be a strong support to strangeon star model~\citep{LGX2013, Guo2014CPC}.

\subsection{Glitches}

Strangeon stars would be solidified during cooling.
The local pressure could be anisotropic in elastic matter, so for solid strangeon stars the radial pressure gradient could be partially balanced by the tangential shear force.
A sudden change of tangential force may result in a starquake, with release of both
gravitational and elastic energy.
The solid stars would be natural to have glitches as the result of starquakes~\citep{Zhou2004,PX2008,Zhou2014,Lai2018MN}.

There are actually two kinds of starquakes in the solid strangeon star model: bulk-invariable (Type I) and bulk-variable (Type II) starquakes, with energy release negligible for the former but significant for the latter.
Both types of glitches with and without X-ray enhancement could be naturally understood in the starquake model of solid strangeon stars~\citep{Zhou2014}.
Moreover, the type II giant starquakes of strangeon stars could reproduce the typical energy of $10^{44-47}$ erg released during superflares of soft gamma-ray repeaters (SGRs)~\citep{XTY2006}.

Dividing the inner motion of the star during starquakes into plastic flow and
elastic motion, the behaviors of glitches without significant energy
releasing (including the Crab and the Vela pulsars), could be understood in an uniform model~\citep{Lai2018MN}.
The plastic flow which is induced in the fractured part of the outer layer, would move tangentially to redistribute the matter of the star and would be hard to recover.
The elastic motion, on the other hand, changes its shape and would recover significantly.
The results under this scenario show consistency with observations, including the recovery coefficient as a function of glitch size, and the time interval between two successive glitches as the function of the released stress.

The glitch activity, i.e. the statistical feature of glitches, of normal radio pulsars could also be reproduced in the framework of starquake of solid strangeon stars~\citep{WangWH2021}.
The fact that typical glitch sizes of two rapidly evolving
pulsars, the Crab pulsar and PSR B0540-69, are about two orders of magnitude lower than that of the Vela pulsar, may indicates that the changes of oblateness during glitches depend on the age of pulsars.
The small glitch sizes and low glitch activity of the Crab pulsar can be explained simultaneously in this phenomenological model, and the energy releases for typical glitch sizes are also obtained.

Besides glitches, solid strangeon stars have some other consequences.
Significant stellar oscillations could usually accompany starquakes, and the magnetospheric activities in the polar cap region of pulsars could be excited under such oscillations, which would influence particle acceleration~\citep{Lin2015}.
The onset of radio emission after glitches/flares in SGRs or AXPs (anomalous X-ray pulsars) could be the result of oscillation-driven magnetic activities of solid strangeon stars.
It is suggested that no gravitational wave originated from r-mode instability could be detected for solid strangeon stars~\citep{Xu2006}.

\subsection{Tidally deformability}

The gravitational wave (GW) event GW170817~\citep{GW170817} and its multiwavelength electromagnetic counterparts (e.g.,~\cite{Kasliwal2017}) open a new era in which the nature of pulsar-like compact stars could be crucially tested.
The tidal deformability indicated by the signals of gravitational waves from binary merger could put a clean and strong constraint on the EOS of compact stars.
In addition, the quickly developed field of fast radio burst (FRB) would also hint the existence of strangeon star, from the phenomenology of either time-frequency drifting~\citep{Wang2019ApJ} or multi-origins~\citep{Jiang2020RAA} of FRBs.

During the late inspiral stage of merging binary compact stars, mass quadrupole moment will be induced by the external tidal field of the companion, which accelerates the coalescence and hence detectable by GW observations~\citep{Flanagan2008}.
This property of compact stars can be characterized by the dimensionless tidal deformability (defined as $\Lambda = (2/3)k_2/(GM/c^2R)^5$ where $k_2$ is the second tidal love number), which depends on EOS.
Different types of EOS predict different values of $\Lambda$, and the constraint of $\Lambda$ by GW 170817~\citep{GW170817} puts robust constraint on EOS of compact stars.

Strangeon stars are self-bound and consequently have finite surface density, which leads to a correction to calculate the tidal deformability.
As a result, they can reach a much higher maximum mass under the same tidal deformability constraint.
We find that, in a quit large parameter-space, strangeon star model survives the scrutiny of GW170817~\citep{Lai2018RAA,LZX2019}, since the tidal deformability of a strangeon star with mass $M=1.4M_\odot$  could be as low as $\Lambda \sim 200$ and the maximum mass $M_{\rm max}$ inferred by its EOS would reach $\sim 3M_\odot$.
By contrast, it is not so easy for neutron star models to pass all the tests.
Neutron stars are gravity-bound, which means that if a neutron star has the same mass as that of a strangeon star, the former usually has a larger radius than that of the latter.
Explicitly, it is found that~\citep{Annala2017}, under all of the possible EOSs of neutron stars, the maximum mass of neutron stars cannot reach higher than $2.8 M_\odot$ in order to satisfy the constraint of tidal deformability given by GW170817.

Recently, the static, the slowly rotating, and the tidally deformed strangeon stars have been investigated in the perturbative approaches of general relativity~\citep{GaoY2022}.
The calculated moment of inertia, $I$, of strangeon stars with mass $\sim 1.4M_\odot$ and spin frequencies beween ($200-600$) Hz is around $2\times 10^{45}$ g$\cdot$cm$^2$, which is consistent with the observed value of the double pulsar PSR J0737-3039A/B~\citep{Kramer2021}.
For an initial spin period $\sim 10$ ms after core-collapse supernova, the moment of inertia increases $\delta I\sim 10^{-3} I$ relative to its non-rotating value.
This is sufficient for star-quake-induced glitches of strangeon stars~\citep{WangWH2021}.

\subsection{Strangeon kilonova}
\label{sec:kilonova}

The multiwavelength electromagnetic counterparts of GW170817, named kilonova, can also give constraints on the nature of compact stars, although not so directly as that given by tidal deformability.
Therefore, it is necessary to explore the kilonova-like event associated with merging strangeon stars.

In the regime of binary neutron star merger, the outer parts of neutron stars are supposed to be made of neutron-rich nuclei as well as neutrons, so merging binary neutron stars would produce neutron-rich ejecta in which r-process nucleosynthesis will happen.
The kilonova in this case could be called neutron kilonova.
In the regime of binary strangeon star merger, the ejected strangeon nuggets would evaporate strangeons, neutrons and protons, and the ejector would eventually be neutron-rich in the equatorial plan by the decay of strangeons.
The binary strangeon star merger would then also give rise to kilonova-like event, called strangeon kilonova~\citep{Lai2018RAA}.

Because of the high maximum mass ($M_{\rm max} \sim 3 M_\odot$), the merger remnant of binary strangeon stars would usually be very long-lived and even to be stable (if $M < M_{\rm max}$), rather than collapsing quickly to a black hole.
Because the emission rates of different particles depend on
temperature, the ejecta could end up with two components, with high and low opacity respectively.
Consequently, even if the total ejected mass would be $\sim 10^{-3} M_{\odot}$ only~\citep{Bauswein2010,ZhuZY2021}, the spin-down power of the long-lived remnant would account for the whole emission of kilonova AT2017gfo associated with GW170817~\citep{Lai2021RAA}.
The strangeon kilonova would have the similar light curves, as well as similar nucleosynthesis, as that of neutron kilonova, passing the test of kilonova observations of merging compact objects.

We could make a comparison between neutron kilonova and strangeon kilonova from the flavour-symmetric point of view, indicated in Fig.\ref{fig:triangle}.
The neutron kilonova scenario in merging binary neutron star is for changing 2-flavoured asymmetry to almost 2-flavoured symmetry, by the reverse mode of neutronization, or simply called inverse-neutronization ($\bar{\mathcal{N}}$).
The strangeon kilonova scenario in merging strangeon star, however, is for changing 3-flavoured symmetry to almost 2- flavoured symmetry, by the reverse mode of strangeonization, or simply called inverse-strangeonization ($\bar{\mathcal{S}}$).
The detailed picture of merging double strangeon stars is expected to be tested by future numerical simulations.

\section{Dark matter}

It is still a challenge to know the nature of dark matter, and a general view point is that dark matter represents a glimpse of ``new'' physics beyond the standard model.
However, the candidate of dark matter in the standard model of particle physics does exist~\citep{Witten1984}.
In addition, as for dark matter candidates without new physics, the primordial black holes (PBHs) is also worth noting~\citep{Carr2016}.
It is found that all the dark matter could be PBHs, and
the possibility that the dark matter is in intermediatemass PBHs of $1 \sim 10^3 M_\odot$ is of special interest in view of the detection of black-hole mergers by LIGO.
With the mass function estimated from the merger rate of LIGO/Virgo O1 and O2 events, the induced GWs from such a curvature perturbation with a Gaussian narrow peak at some small scale would be in a seemingly mild tension with current constraints from pulsar timing array (PTA)~\citep{Cai2019}, but besides this model-dependent argument, PBHs as dark matter are nevertheless attractive for gravitational wave sciences (e.g., the Advanced Virgo).
Here we focus on the relics of cosmological QCD phase transition as the dark matter candidate~\citep{Witten1984}.

Strange nuggets would form during the cosmological QCD phase transition if the transition is of the first order, and the nuggets with initial baryon number $A > 10^{40}$~\citep{Bhattacharjee1993} would survive after boiling and evaporation, manifested in the form of invisible ``dark matter''.
Different from the weakly interacting massive particles (WIMPs), strange nuggets as dark matter could give a natural explanation to the comparable ratio of dark matter to normal baryonic matter, $\sim$ 5 : 1 (only order of one)~\citep{Witten1984}.
Within this framework, the model of anti-quark matter nuggets~\citep{Zhitnitsky2003,Ge2018} was also proposed for the dark matter candidate.
This model could simultaneously explain the observed relation $\Omega_{\rm DM} \sim \Omega_{\rm visible}$ and the observed asymmetry between matter and antimatter in the Universe (known as the ``baryogenesis'' problem), which two seemly unrelated problems in cosmology.
The axion domain walls were also introduced so that quark matter and anti-quark matter nuggets would form even if the QCD phase transition is of second order or a crossover.

If the formation of nuggets was possible and the resultant nuggets would survive after boiling and evaporation, they would manifest in the form of invisible ``dark matter'', although the quantitative results are related to the explicit forms of EOS of strange quark matter.
Although most of the studies are based on the strange quark matter (i.e. almost free quarks) and the quantitative results do depend on the explicit forms of EOS, the formation of nuggets would still be possible if quarks inside strong matter are localized in strangeons.
In such cases, the resultant strangeon nuggets would have interesting consequences.
The black hole formed by assembling the dark matter nuggets, before the formation of the primodial stars, could be the seeds for the supermassive black holes ($M\sim 10^9M_{\odot}$) at redshift as high as $z\sim 6$~\citep{LX2010}.
In our Galaxy, the accretion of the dark matter nuggets by pulsars would lead to small glitches whose glitch sizes are smaller than those predicted in the standard starquake model~\citep{Lai2016RAA}.

Besides, strangeon nuggets dark matter is naturally self-interaction, so it could provide a useful way for us to understand observations.
For example, some observations find that dark matter halos of some dwarf galaxies are less dense in their central regions compared to expectations from collision-less N-body simulations.
To fit the rotation curves of a sample of galaxies in a self-interacting dark matter model, a velocity-dependent value of $\sigma/m \sim (0.1 \sim 3) \rm cm^2 /g$~\citep{Kaplinghat2016,Kamada2017} is needed, where $\sigma$ is the scattering cross section and m is the dark matter particle mass.
For a charged strangeon nugget with baryon number $A = A_{30}10^{30}$, the mass is $m \simeq 1.6 \times 10^6A_{30}$ g, and the Debye length is $\lambda_{\rm D} = \sqrt{kT/(4\pi ne^2)}$ with $T$ the temperature and $n$ the number density of free particles charged, at the scale of which the nugget's electric fields are screened in interstellar medium.
Then the value of $\sigma/m$ for self-interacting strangeon dark matter can be estimated as
\begin{equation}
    (\sigma/m)_{\rm strangeon\ dark\ matter}\simeq \lambda_{\rm D}^2/m \sim (3\ {\rm cm^2/g})\cdot T_5n_1A_{30}^{-1},
\end{equation}
where $T = T_5 \times 10^5$ K and $n = n_1 \times 1\ \rm cm^{-3}$.
This value would be within the allowed range inferred by the observations.

Non-relativistic strangeon nugget with $A \sim 10^{30}$ could be the cold dark matter candidates, but conventional experiments for directly detection of dark matter particles are not sensitive to strangeon dark matter.
If current detectors with increasing sensitivity would further fail to catch the elusive dark matter particles, strangeon nuggets could still be considered as the candidate of dark matter.
Certainly, observational tests of the strange nugget (either quark nugget or strangeon nugget) as dark matter candidates are surely surely important.
Several exotic events reported by the Pierre Auger Observatory could be explained within the so-called axion quark nugget dark matter model~\citep{Zhitnitsky2022}, but the interaction between the nugget and its propagating matter would be enhanced if it is strongly magnetized, resulting in a magnetopause with surrounding plasma~\citep{VanDevender2021a,VanDevender2021b}.
It is then very necessary to model the interaction details in order to identify a quark nugget or a strangeon nugget, to be meaningful not only in dark matter detection, but also in the research of cosmic rays noted as following.

\section{Cosmic rays}

Stable strangeon nuggets with baryon number $A$ larger than the critical value  $A_{\rm c}$ could be ejected relativistically or non-relativistically in the binary strangeon star merger.
In the strangeon kilonova event of merging strangeon stars, although the ejected strangeon nuggets would suffer from evaporation, a significant amount of strangeon nuggets would survive~\citep{Lai2021RAA} with baryon number $A>A_{\rm c}$.
These survived strangeon nuggets may fly away and reach the Earth through long-time travel, and eventually result in a strangeon nuggets cosmic ray air-shower in the Earth's atmosphere.
The strangeon nuggets cosmic rays usually have high energy.
For instance, the rest mass of a nugget with $A \sim 10^{10}$ is $\sim 10^{19}$ eV, and the deposit energy during corresponding air shower could then be of order $10^{18\sim 20}$ eV, depending certainly on its speed.
This kind of strangeon nuggets cosmic rays remains a possibility up-to-now, for the existing experiments are only sensitive to nuggets with $A < 10^5$.
It is also found that, the small lumps of strange quark nuggets (strangelets) as ultra-high energy cosmic ray can circumvent the so-called GZK-cutoff problem~\citep{Madsen2003}.

We can briefly discuss the air-shower of strangeon nuggets cosmic rays.
For Lorentz factor $\gamma <2$, the kinematic energy of a strangeon nugget cosmic ray is $E_{\rm CR} \sim m_{\rm CR}c^2\beta^2$, where $\beta = \sqrt{1-\gamma^{-2}}$ measures the speed and $m_{CR}$ is the rest mass.
We can then have
\begin{equation}
    E_{\rm CR} \sim (10^{17}{\rm eV}) A_{10}\beta^2_{0.1},
\end{equation}
where the baryon number $A = A_{10} \times 10^{10}$ and $\beta = 0.1\beta_{0.1}$.
Therefore, in the cosmic ray rest frame, a proton in Earth's atmosphere has a kinematic energy of
\begin{equation}
    E_{\rm proton} \sim (10{\rm MeV}) \beta^2_{0.1}.
\end{equation}
Because the hadronic cascade may stop when $E_{\rm proton} < m_{ \pi}c^2 \sim 100$ MeV (or $\beta < 0.3$), and the electromagnetic cascade may end when $E_{\rm proton} < 2m_{\rm e}c^2 \sim 1$ MeV (or $\beta < 0.03$), the interaction should become very weak if the speed of the strangeon nugget is lower than $\sim 10^9{\rm cm/s}$, when the nugget would go almost freely through the Earth.
On the other hand, in case of $\beta > 0.1$, an atomic nucleus would be destroyed during collision, since the nuclear binding energy per baryon is comparable to $E_{\rm proton}$.
Assuming PeV-energy deposit in air-shower and $\sim$ 100 MeV-energy lose per nucleon during interaction, we may estimate the atmospheric depth, $X$, by $10^8 {\rm eV}\cdot (X/m_{\rm p}) \cdot (A^{1/3} {\rm fm})^2 \sim 10^{15}$eV,
\begin{equation}
    X \sim 10^7A^{-2/3}m_{\rm p} \sim (400 {\rm g/cm^2}) \cdot A_{10}^{-2/3}.
\end{equation}
Without doubt, it is worth waiting for an identification of strangeon nuggets cosmic rays either in low altitude (e.g., the Pierre Auger Observatory) or in high altitude (e.g., the Large High Altitude Air Shower Observatory, abbreviated to LHAASO), or even extra-ordinary detectors beyond conventional imaginations.

\section{Summary and Outlook}

Nature loves symmetry, although we usually find symmetry breaking at negligible scale.
Heavy flavours of quarks will not participate in bulk strong matter at a few nuclear matter densities involved in astrophysics, so the {\em principle of flavour maximization}~\citep{Xu2018} is limited only to light flavour symmetry: within either 2-flavours or 3-flavours.
The coupling between quarks is so strong that quarks are localized either in 2-flavoured nucleons or 3-flavoured strangeons.
Even Nature loves the principle of flavour maximization, microscopic strong matter, i.e. normal nuclei, could be only 2-flavoured as the weak interaction can easily convert $s$-quark to $u/d$-quark, producing electrons with negligible kinematic energy.
However, macroscopic strong matter should be 3-flavoured, otherwise the system would be unstable due to an extremely high energy of electron or a high nuclear symmetry energy.
It would be possible to build a kind of stable bulk strong matter, composed of nucleon-like but 3-flavored units, i.e. the strangeon matter.
3-flavored strangeons, which might constitute macroscopic and even cosmic strong matter, could be manifested as trinity: strangeon stars, strangeon dark matter  nuggets and strangeon cosmic ray nuggets, as focused in this article.

The inner structure of pulsar-like compact objects as well as the EOS of supra-nuclear dense matter are challenging in both physics and astronomy.
In the view of strangeon matter, both perturbative and non-perturbative QCD effects are responsible to solving the problem: the former results in quark-flavour maximization, and the latter results in the localization of
quarks inside strangeons as well as a hard core repulsion between strangeons.
If quarks are localized in strangeons, and additionally a hard core repulsion exists between strangeons due to significant non-perturbative QCD effects, then the EOS of strangeon stars should be very stiff and the energy per baryon could also be the lowest.
The observational consequences of strangeon stars show that different manifestations of pulsar-like compact stars, from the surface and global properties to the bolometric radiation of associated kilonova of merging binaries, could be understood in the regime of strangeon stars.

Bulk strong matter could actually be strangeon matter, and could be manifested in the form of compact stars, dark matter, and cosmic rays, as explained in this review.
It is expected that, advanced facilities will identify the existence of strangeon matter, particularly in the new era of multi-messenger astronomy.

\begin{acknowledgments}
This work is supported by the National SKA Program of China (Nos. 2020SKA0120300, 2020SKA0120100), the
National Key R\&D program of China (No. 2017YFA0402602), the National Natural Science Foundation of China
(Grant Nos. U1831104, 12275234), and the Outstanding Young and Middle-aged Science and Technology Innovation
Teams from Hubei colleges and universities (No. T2021026).
\end{acknowledgments}


\end{document}